\documentclass[11pt]{article}
\usepackage[margin=1in]{geometry}
\usepackage[T1]{fontenc}
\usepackage{lmodern}
\usepackage{booktabs}
\usepackage{graphicx}
\usepackage{amsmath}
\usepackage{array}
\usepackage[numbers,sort&compress]{natbib}
\usepackage[hidelinks]{hyperref}
\hypersetup{pdftitle={Escalate on Failure: A Single-Model Mode-Switching Cascade for Local Code Generation},pdfauthor={Brian K. Hudes, MD}}
\makeatletter\g@addto@macro\UrlBreaks{\do\-}\makeatother

\title{Escalate on Failure: A Single-Model Mode-Switching Cascade\\for Local Code Generation}
\author{Brian K. Hudes, MD\\AiMOS Systems Corporation}
\date{November 2026}

\begin{document}
\maketitle

\begin{abstract}
\noindent Preliminary report. We study escalation on failure for code generation with locally served open-weight models: one coder model tries fast, and when a test run fails the same model is switched to reasoning mode. On the 225 exercises of the public Aider Polyglot benchmark, each exercise gets a fixed schedule of tries and stops at the first passing test run. In-sample (stage boundaries were tuned on the same exercises), every escalation run solved 215 to 222, against 205 (91.1\%) for a settings-matched baseline of ten fast tries. A second model was not needed to reach these solve counts: the two-model cascade (second model for the last two tries) solved 219 in each of two runs. The data neither show nor exclude a small benefit from the second model. At 4+4 (n+m: n fast tries, then m reasoning tries), starting the reasoning tries in a fresh conversation solved 222 and 219, against 215 to 217 with the conversation kept, a likely modest benefit of unestablished size; at 2+4 the comparison is mixed. Escalating earlier (after two or three fast tries) gave no accuracy gain distinguishable from run-to-run noise and took longer: 2+4 solved 220 and 216, 3+4 solved 219 and 221, against 216, 215 and 217 for 4+4. Run-to-run noise is a few exercises (identical 2+4 runs solved 220 and 216; identical 4+4 runs differed on nine exercises), so small differences are not established. The gain is more exercises solved, not speed: the median time to a passing solution is not lower (10.3\,s for the baseline, 10.9 to 14.8\,s for escalation runs), and wall clock is longer (2\,h 17\,min to 5\,h 00\,min against 2\,h 07\,min). These in-sample work-until-correct figures are not comparable with two-try leaderboard scores. Other limits: possible training-data overlap, a single baseline run and one to three runs per variant.
\end{abstract}

\section{Introduction}
Work on local inference for code generation is usually framed around a single model: which model, which quantisation, which engine, how many tokens per second, with recent position work arguing instead that small models, possibly combined, suffice for many agentic tasks \citep{belcak2025slm}. A practitioner who wants working code is asking a different question, namely how long it takes to obtain an answer that passes the tests. That time depends on the model, but also on how failures are handled: how many retries are allowed, whether the model changes its way of working after a failure, and whether a second model is brought in.

This report describes escalation on failure on one workstation with two GPUs. A fast coder model tries first; each failing test run triggers another try with the test output; after a fixed number of fast tries (four in the main configuration) the same model is switched to reasoning mode (called ``thinking'' in the tables and engine settings) for further tries. The schedule stops at the first passing test run. We call this work-until-correct: a schedule of tries that stops at the first passing test run, in contrast with the leaderboard's two-try rule (an initial try plus one retry with test output). The central claim concerns this switch of one model's mode. We also tested a two-model cascade, in which a second model with reasoning takes over for the last two tries. Below, ``the two-model cascade'' denotes that schedule, and the single-model schedules are named by their numbers of fast and reasoning tries (for example 4+4). ``Cascade'' alone means the two-model system; the single-model version is called a schedule or single-model escalation, and the title's use of ``cascade'' is the generic sense. ``Keep'' and ``fresh'' say whether the conversation is kept or reset at the switch from fast to reasoning tries. The ablations indicate that most of the gain comes from the reasoning-mode switch within one model; the second model was tested, was not needed to reach the solve counts observed, and the data neither show nor exclude a small benefit from it. We evaluate the schedules on the Aider Polyglot benchmark against a settings-matched single-model baseline, and report wall-clock time to a passing solution, not only the share of exercises solved. Settings-matched means the baseline uses the same harness, summaries off (aider's chat-history summaries, which condense earlier turns, are disabled), one thread and the same coder server; the two-model cascade additionally has a second model on a second GPU and a lower try cap (8 versus 10), and the single-model schedules use 6 to 8 tries. In short: after four fast tries, switching the same model into reasoning mode solved 10 to 17 more of the 225 exercises than ten fast tries did, with no gain in speed.

The contributions are modest and are stated as such.
\begin{enumerate}
\item A single-model mode-switching schedule: four fast tries, then four tries with the same model in reasoning mode, on one GPU. Three runs solved 216, 215 and 217 of 225 exercises (the third with a time cap that never fired), against 205 for a settings-matched baseline with ten fast tries.
\item Eleven further single-model runs: a repeat of 4+4, the time-capped run, escalation after three and after two fast tries (two runs each), and a fresh conversation at the mode switch (4+4: two corrected runs and one provisional run; 2+4: two runs). All eleven reached 215 to 222 and all beat the baseline in paired counts. At 4+4, a fresh conversation at the switch is consistent in direction across the runs and a likely modest benefit, but its size is not established (at 2+4 the comparison is mixed); earlier escalation gave no accuracy gain distinguishable from run-to-run noise and took longer; differences of a few exercises between runs are within run-to-run noise.
\item A two-model cascade (four fast tries, two reasoning tries with the same model, two tries with a second model), run twice unchanged, with both runs solving 219 of 225 exercises. A second model was not needed to reach the solve counts observed; the data neither show nor exclude a small benefit from it.
\item A per-attempt yield analysis showing how quickly the success rate of repeated fast tries decays, and how much of it reasoning mode restores.
\item Time to a passing solution, rather than try count, as the reported metric. We report it although escalation does not win on it: the matched baseline has the lowest median. The result is more exercises solved, not a lower median time.
\item A set of negative results (larger model, always-on reasoning, concurrent requests, prompt caching, routing by language).
\end{enumerate}
We make no claim to beat hosted frontier models. The evidence comes from fifteen runs: the matched baseline once, 4+4 keep three times, 4+4 fresh twice with the corrected note (and once provisionally), 2+4 keep twice, 2+4 fresh twice, 3+4 keep twice and the two-model cascade twice, so effect sizes below a few exercises cannot be resolved (Section~\ref{sec:limits}).

\section{Related work}
\label{sec:related}
\paragraph{Cascades and routing.}
Cost-oriented cascades send a query to cheaper models first and escalate when the answer looks unreliable. FrugalGPT learns such a cascade over commercial APIs \citep{chen2023frugalgpt}. AutoMix has a small model check its own answer and escalates through a learned router \citep{aggarwal2023automix}. RouteLLM and Hybrid LLM instead train routers that choose between a strong and a weak model before generation \citep{ong2024routellm,ding2024hybrid}. Dekoninck et al.\ analyse routing and cascading in one framework and show that the quality of the estimator driving the decision is what matters most \citep{dekoninck2024unified}; a test run is a direct estimator for code. A recent survey covers the area \citep{moslem2026survey}. Closest in spirit are SWE-Router, in which a cheap model works for a few turns and a router inspects the partial trajectory before deciding to escalate \citep{son2026swerouter}, and R2V, in which a small model is paired with a verifier and a calibrated router that calls a teacher model only when failure is likely \citep{hemadri2026r2v}. These optimise cost or escalation rate, use a learned or calibrated decision signal, and, to our knowledge, do not report wall-clock time to a passing solution on a locally served model pair. Here the signal is simply a failing test run, and the metric is time.

\paragraph{Ensembles, mixtures and debate.}
Mixture-of-Agents has layers of models refine each other's outputs, so every model runs on every query \citep{wang2024moa}. LLM-Blender ranks candidates pairwise and fuses them \citep{jiang2023llmblender}. Repeated sampling from one model with majority voting improves with the number of samples \citep{li2024moreagents}, and multi-agent debate has instances critique each other over several rounds \citep{du2023debate}. These approaches improve answer quality by combining outputs, usually without an executable check. The cascade here never runs a later stage unless an earlier one has failed a test. Vallecillos-Ruiz et al.\ measure the complementarity of ten models on code generation and repair and study how to select among their outputs \citep{vallecillos2025wisdom}; their setting is offline selection, whereas ours is online escalation with a test oracle.

\paragraph{Small models and compound systems.}
Belcak et al.\ argue in a position paper that small language models are sufficiently capable, and more economical, for many agentic tasks \citep{belcak2025slm}. This work is an empirical data point on one test-verified coding task, not a position.

\paragraph{Generate-and-verify and self-repair.}
AlphaCode samples very many programs and filters them by behaviour on tests \citep{li2022alphacode}. CodeT has a model generate tests for its own candidates and selects by agreement \citep{chen2022codet}. Reflexion lets an agent reflect verbally on execution feedback and retry \citep{shinn2023reflexion}. Olausson et al.\ find that gains from self-repair are small once the cost of repair is counted, and that stronger feedback helps more \citep{olausson2023selfrepair}. The two closest reports to this work are small-scale local studies: one pairs a code-writing model with a reasoning reviewer on modest hardware and evaluates on HumanEval+ and MBPP+ \citep{miller2026review}, and another finds that execution feedback mattered more than added pipeline topology in pipelines of 1--3B-parameter models \citep{mcandrews2026feedback}. Both evaluate on HumanEval- and MBPP-family benchmarks rather than Aider Polyglot. On the benchmark used here, aider's architect mode pairs a reasoning model with an editing model in sequence \citep{aider2024architect}. What differs in our setting is the combination of an open-weight, locally served model pair, escalation triggered by failing tests, mode switching inside one model, and time to a passing solution as the headline measure. Each ingredient is known; we do not claim the idea of escalation as new.

\section{Method: escalation schedules}
\label{sec:method}
The main schedule is a fixed sequence of up to eight tries per exercise with one model, four fast tries followed by four tries in reasoning mode (4+4). Variants change the number of fast tries (3+4 and 2+4), limit the duration of reasoning requests (a 300\,s cap), or reset the conversation at the switch (``fresh''; in the other runs the conversation is kept, ``keep''). The two-model cascade replaces the last two reasoning tries with a second model (Table~\ref{tab:cascade}). Every try after the first receives the failing test output, as in the standard benchmark. A schedule stops at the first passing test run. The model proposes edits, the harness applies them and runs the exercise's test suite, and the test result alone decides whether to continue. There is no learned router and no model-based judge.

\begin{table}[t]
\centering
\caption{The two-model cascade. Both models are resident for the whole run, one per GPU. The single-model 4+4 schedule uses the first two rows' model and settings for tries 1--4 and thinking on for tries 5--8, with no second model.}
\label{tab:cascade}
\begin{tabular}{llllll}
\toprule
Tries & Model & Mode & Engine & Placement \\
\midrule
1--4 & Qwen3.8-Flash-Next Q8\_0 & thinking off & Basalt v0.5.5 & GPU 0 + RAM \\
5--6 & Qwen3.8-Flash-Next Q8\_0 & thinking on & Basalt v0.5.5 & GPU 0 + RAM \\
7--8 & DeepSeek V4.1-Flash 2-bit & thinking on & Maya & GPU 1 \\
\bottomrule
\end{tabular}
\end{table}

\paragraph{Ordering.}
The order follows from earlier single-model runs. Flash-Next with thinking off had the shortest time to a correct answer of anything tested among models with usable accuracy (median about 12\,s). With thinking always on, the same model scored 85.8\% on two tries, but the median rose to 2.5\,min and the run took 14\,h 22\,min. Reasoning is worth its cost only on the exercises that fast mode cannot solve, so it is placed behind four fast tries.

\paragraph{Hand-off rules.}
\begin{itemize}
\item \emph{Fast to thinking (same model).} In the main schedule the chat is kept (``keep''); the thinking tries see the earlier attempts and their test output. In the ``fresh'' variants the conversation is reset at the switch and the model is given a short brief stating that the earlier attempts were its own quick attempts and did not pass the tests (Section~\ref{sec:fresh}). One earlier fresh run used a brief that wrongly said the earlier attempts were by another model; it is reported as provisional and kept separate. Harness version 0.3.1 was used for the six further runs; it differs from 0.3.0 only in the wording of this brief.
\item \emph{Flash-Next to DeepSeek.} A fresh chat is started. DeepSeek receives a brief containing the original task, a one-line list of earlier attempts and the latest test output. The solution files are passed in their current state.
\item \emph{Sampling.} Temperature is 0 for both models in every stage (the harness default). No top\_p is sent, so each server's own value applies (0.95 on Basalt, 1.0 on Maya).
\item \emph{Timeouts.} Request timeouts are 600\,s for fast tries, 1800\,s for Flash-Next thinking tries and 3600\,s for DeepSeek. There is no per-exercise time cap beyond the eight-try limit and a three-minute test timeout. In one variant the timeout for thinking requests was lowered to 300\,s; no request reached it (Section~\ref{sec:timecap}), so the cap is untested.
\end{itemize}

\section{Experimental setup}
\label{sec:setup}
Table~\ref{tab:setup} lists the configuration. The benchmark is the Aider Polyglot set of 225 Exercism exercises. Both models run on a single workstation and nothing leaves the machine. The engines and models are described in the results repository.

\begin{table}[t]
\centering
\caption{Experimental setup.}
\label{tab:setup}
\begin{tabular}{@{}lp{4.9in}@{}}
\toprule
Item & Value \\
\midrule
Benchmark & Aider Polyglot, 225 Exercism exercises (C++ 26, Go 39, Java 47, JavaScript 49, Python 34, Rust 30) \\
Harness & aider commit \texttt{5dc9490}, \texttt{diff} edit format, 1 benchmark thread \\
Harness change & Chat-history summaries off (Section~\ref{sec:limits}) \\
Harness version & Harness 0.2.2 for the earlier runs, 0.3.0 for the five earlier single-model variants (behaviour identical with the new option off), 0.3.1 for the six further runs (differs from 0.3.0 only in the wording of the fresh-conversation hand-off brief) \\
Machine & Dell Precision T7920, dual Xeon Platinum 8168, 512\,GB DDR4 ECC \\
GPUs & 2$\times$ NVIDIA RTX PRO 6000 Blackwell Max-Q, 96\,GB each, PCIe 3.0 \\
Coder & Qwen3.8-Flash-Next, Unsloth Q8\_0, as a 179\,GiB \texttt{.basalt} pack on Basalt v0.5.5, GPU 0 plus system RAM, 64k context \\
Second model & DeepSeek-V4.1-Flash-1017, Maya-S IQ2\_XXS (about 2-bit), on the Maya engine, GPU 1 \\
\bottomrule
\end{tabular}
\end{table}

\paragraph{Measured speeds.}
From earlier single-request measurements on this hardware: Flash-Next on Basalt reads prompts at about 1.36k tok/s and decodes at 132 to 141 tok/s, or 238 to 328 tok/s with speculative decoding (8-token drafts, 74 to 92\% accepted). DeepSeek at 2-bit on Maya decodes at about 35 tok/s on one card and about 72 tok/s across two.

\paragraph{Comparison runs.}
The matched baseline is Flash-Next alone, thinking off, up to ten tries, with the two-model cascade's settings (summaries off, one thread, same harness, same Basalt server). An ablation run uses the same settings with four fast tries and then four thinking tries from Flash-Next and no second model. The two-model cascade was also run a second time with nothing changed. Eleven further single-model runs (Section~\ref{sec:variants}) used the same settings; two model instances ran side by side, one per GPU, so wall clocks of concurrently running runs are not strictly comparable.

\section{Results}
\label{sec:results}
The single-model schedules are reported against the matched baseline first (Sections~\ref{sec:variants} to~\ref{sec:timecap}). The two-model cascade follows as one of the configurations tested, an ablation that adds a second model (Section~\ref{sec:cascade}). Schedules are named by their numbers of fast and reasoning tries (``4+4'' is four fast tries, then four reasoning tries with the same model); ``keep'' and ``fresh'' say whether the conversation is kept or reset at the switch from fast to reasoning tries (Section~\ref{sec:method}).

\subsection{Single-model escalation against the matched baseline}
\label{sec:variants}
Twelve single-model runs used the same settings as the matched baseline (Flash-Next alone, summaries off, one thread, temperature 0, one GPU for the model, no second model). They cover five schedules: 4+4 keep (three runs, one of them with a 300\,s limit on thinking requests that never fired), 3+4 keep (two runs), 2+4 keep (two runs), 4+4 fresh (two runs with a corrected hand-off note, plus one earlier \emph{provisional} run whose note wrongly said that the earlier attempts were by another model) and 2+4 fresh (two runs). Runs of one schedule are numbered in the order they were run. The eleven runs after the first 4+4 run all completed 225 of 225 exercises with no infrastructure errors, and no try hit a model timeout. Table~\ref{tab:variants} gives the keep schedules and Table~\ref{tab:variantsfresh} the fresh ones, each with the matched baseline for reference; Table~\ref{tab:vpaired} gives the paired counts against the baseline. The provisional run is set apart in every table and is not pooled with the corrected runs anywhere in this report.

\begin{table}[t]
\centering
\caption{Single-model schedules with the conversation kept at the switch, with the matched baseline and the two-model cascade for reference. ``Solved at try'' lists exercises solved at each try, fast tries before the slash and thinking tries after it (baseline: tries 1--8; tries 9--10 solved none). ``Within 2 tries'' is the share of the 225 exercises solved by try 2 (the leaderboard-style score); ``Median'' is the median time to a passing solution among solved exercises. Wall clocks of runs that ran side by side are not strictly comparable. The 300\,s cap never fired, so that run is in effect a third repeat of 4+4.}
\label{tab:variants}
\footnotesize
\setlength{\tabcolsep}{3pt}
\begin{tabular}{@{}lccccl@{}}
\toprule
Schedule & Solved & Within 2 tries & Median & Wall clock & Solved at try \\
\midrule
Matched baseline, 10 fast & 205 (91.1\%) & 76.9\% & 10.3\,s & 2\,h 07\,min & 90, 83, 23, 2, 4, 2, 0, 1 \\
\midrule
4+4 keep, first run & 216 (96.0\%) & 75.1\% & 10.9\,s & 2\,h 33\,min & 85, 84, 22, 10 / 9, 5, 1, 0 \\
4+4 keep, repeat & 215 (95.6\%) & 74.7\% & 12.1\,s & 2\,h 44\,min & 82, 86, 27, 4 / 12, 4, 0, 0 \\
4+4 keep, 300\,s cap (inert) & 217 (96.4\%) & 75.6\% & 12.5\,s & 2\,h 17\,min & 88, 82, 25, 6 / 13, 1, 2, 0 \\
3+4 keep, first run & 219 (97.3\%) & 72.4\% & 12.8\,s & 3\,h 15\,min & 81, 82, 26 / 19, 6, 3, 2 \\
3+4 keep, second run & 221 (98.2\%) & 73.3\% & 12.1\,s & 3\,h 03\,min & 89, 76, 27 / 18, 7, 3, 1 \\
2+4 keep, first run & 220 (97.8\%) & 75.1\% & 12.8\,s & 3\,h 11\,min & 84, 85 / 38, 10, 1, 2 \\
2+4 keep, second run & 216 (96.0\%) & 71.1\% & 13.9\,s & 3\,h 16\,min & 84, 76 / 41, 11, 4, 0 \\
\midrule
Two-model cascade, run 1 & 219 (97.3\%) & 75.1\% & 11.5\,s & 3\,h 39\,min & 85, 84, 21, 7 / 13, 7 / 1, 1 \\
Two-model cascade, run 2 & 219 (97.3\%) & 71.6\% & 12.8\,s & 3\,h 26\,min & 83, 78, 31, 8 / 12, 2 / 3, 2 \\
\bottomrule
\end{tabular}
\end{table}

\begin{table}[t]
\centering
\caption{Single-model schedules with a fresh conversation at the switch, with the matched baseline for reference; columns as in Table~\ref{tab:variants}. Runs 1 and 2 used the corrected hand-off note. The run below the rule used a note that wrongly said the earlier attempts were by another model; it is provisional and is not pooled with the others. The 2+4 fresh runs ran about 5\,h, longer than any other escalation run (Section~\ref{sec:earlier}).}
\label{tab:variantsfresh}
\footnotesize
\setlength{\tabcolsep}{3pt}
\begin{tabular}{@{}lccccl@{}}
\toprule
Schedule & Solved & Within 2 tries & Median & Wall clock & Solved at try \\
\midrule
Matched baseline, 10 fast & 205 (91.1\%) & 76.9\% & 10.3\,s & 2\,h 07\,min & 90, 83, 23, 2, 4, 2, 0, 1 \\
\midrule
4+4 fresh, run 1 & 222 (98.7\%) & 77.8\% & 13.1\,s & 3\,h 27\,min & 87, 88, 21, 5 / 13, 6, 0, 2 \\
4+4 fresh, run 2 & 219 (97.3\%) & 76.9\% & 11.9\,s & 3\,h 22\,min & 90, 83, 16, 13 / 13, 2, 1, 1 \\
2+4 fresh, run 1 & 222 (98.7\%) & 72.4\% & 14.8\,s & 5\,h 00\,min & 84, 79 / 49, 8, 1, 1 \\
2+4 fresh, run 2 & 219 (97.3\%) & 74.2\% & 12.7\,s & 4\,h 52\,min & 87, 80 / 42, 6, 2, 2 \\
\midrule
4+4 fresh, provisional & 220 (97.8\%) & 79.6\% & 11.5\,s & 3\,h 24\,min & 82, 97, 16, 9 / 11, 3, 2, 0 \\
\bottomrule
\end{tabular}
\end{table}

\begin{table}[t]
\centering
\caption{Paired per-exercise outcomes of the single-model runs against the matched baseline (exercises solved only by the run named / only by the baseline), and, in the last column, against the first 4+4 keep run (only the run named / only the first 4+4 run). The $p$-values are exact McNemar, nominal and uncorrected; each comparison uses single runs. n/c: not computed for the six further runs. No $p$-values are given for comparisons between variants. The run below the rule is provisional.}
\label{tab:vpaired}
\footnotesize
\begin{tabular}{@{}lccc@{}}
\toprule
Run & vs matched baseline & $p$ & vs first 4+4 keep run \\
\midrule
4+4 keep, first run & 15 / 4 & 0.019 & --- \\
4+4 keep, repeat & 12 / 2 & 0.013 & 4 / 5 \\
4+4 keep, 300\,s cap & 13 / 1 & 0.002 & 5 / 4 \\
3+4 keep, first run & 14 / 0 & 0.0001 & 5 / 2 \\
3+4 keep, second run & 17 / 1 & n/c & --- \\
2+4 keep, first run & 17 / 2 & 0.0007 & 4 / 0 \\
2+4 keep, second run & 13 / 2 & n/c & --- \\
4+4 fresh, run 1 & 17 / 0 & n/c & 8 / 2 \\
4+4 fresh, run 2 & 14 / 0 & n/c & 5 / 2 \\
2+4 fresh, run 1 & 17 / 0 & n/c & --- \\
2+4 fresh, run 2 & 15 / 1 & n/c & --- \\
\midrule
4+4 fresh, provisional & 15 / 0 & 0.00006 & 5 / 1 \\
\bottomrule
\end{tabular}
\end{table}

\paragraph{Every escalation run beat the matched baseline.}
In every paired comparison (Table~\ref{tab:vpaired}) the single-model run solved 12 to 17 exercises that the baseline did not, against 0 to 4 the other way round. The solved counts, 215 to 222, all exceed the baseline's 205. For the six runs with a $p$-value in the table, the nominal values run from 0.019 to 0.00006, uncorrected and from single runs; none was computed for the six further runs. This extends the result of Section~\ref{sec:matched} from one single-model run to twelve. Both runs of the two-model cascade also beat the baseline (Section~\ref{sec:cascade}), so all fourteen escalation runs did. Every comparison above rests on a single baseline run (the baseline was run once, so its own run-to-run spread is unmeasured).

\paragraph{Run-to-run noise.}
Runs of the same configuration differ by a few exercises (Table~\ref{tab:repeats}). Two 4+4 keep runs solved 216 and 215, and nine exercises changed outcome between them (4 solved only in the repeat, 5 only in the first run); the run with the inert cap, in effect a third repeat, solved 217. Two 2+4 keep runs solved 220 and 216 (5 only in the first, 1 only in the second), two 3+4 keep runs 219 and 221 (1 only in the first, 3 only in the second), the two corrected 4+4 fresh runs 222 and 219 (4 only in run 1, 1 only in run 2) and the two 2+4 fresh runs 222 and 219 (3 only in run 1, 0 only in run 2). The within-two-tries score ranged from 71.1\% to 79.6\% across all runs, although tries 1 and 2 are configured identically in every run, so that score moves by several points with no change in configuration. The final counts of all fourteen escalation runs (twelve single-model and two two-model) fall between 215 and 222. Differences of a few exercises between variants are therefore not established.

\begin{table}[t]
\centering
\caption{Paired per-exercise outcomes of repeated runs of the same configuration: exercises solved only by the first-named run / only by the second-named run. The cap in the second row never fired.}
\label{tab:repeats}
\footnotesize
\begin{tabular}{@{}lccc@{}}
\toprule
Configuration & First-named run (solved) & Second-named run (solved) & Only first / only second \\
\midrule
4+4 keep & repeat (215) & first run (216) & 4 / 5 \\
4+4 keep & 300\,s cap run (217) & first run (216) & 5 / 4 \\
3+4 keep & second run (221) & first run (219) & 3 / 1 \\
2+4 keep & second run (216) & first run (220) & 1 / 5 \\
4+4 fresh, corrected note & run 1 (222) & run 2 (219) & 4 / 1 \\
2+4 fresh & run 1 (222) & run 2 (219) & 3 / 0 \\
Two-model cascade & run 1 (219) & run 2 (219) & 2 / 2 \\
\bottomrule
\end{tabular}
\end{table}

\subsection{Earlier escalation}
\label{sec:earlier}
Escalating after two or three fast tries gave no accuracy gain that can be separated from run-to-run noise. With the conversation kept, two runs of 2+4 solved 220 and 216 and two runs of 3+4 solved 219 and 221, against 216, 215 and 217 for 4+4. The repeats moved 2+4 from 220 to 216 (paired 1 against 5), and the 2+4 results overlap those of 4+4. 3+4 keep (219, 221) was 2 to 6 exercises above the three 4+4 keep runs (215 to 217), a gap comparable to the 1 to 4 exercises seen between identical runs, though its upper end (6) is larger; 3+4 keep may be slightly better, but this is not established. With a fresh conversation, 2+4 reached 222 and 219, level with the corrected 4+4 fresh runs (222 and 219); the two 222 runs, 4+4 fresh run 1 and 2+4 fresh run 1, differ on 2 exercises each way.

Earlier escalation costs time. With the conversation kept, 2+4 took 3\,h 11\,min and 3\,h 16\,min and 3+4 took 3\,h 15\,min and 3\,h 03\,min, against 2\,h 17\,min to 2\,h 44\,min for 4+4, because more exercises reach the slow reasoning tries when fewer fast tries come first: 56 and 65 exercises were still open at the switch in the 2+4 keep runs and 36 and 33 in the 3+4 keep runs, against 24 to 26 in the 4+4 keep runs (Table~\ref{tab:rescued}). The two 2+4 fresh runs were the slowest of all the escalation runs: 5\,h 00\,min and 4\,h 52\,min against 3\,h 27\,min and 3\,h 22\,min for the corrected 4+4 fresh runs, about 45\% more wall clock for the same accuracy. Their thinking tries produced about 2.0 million completion tokens each (2,029,976 and 1,999,793), against about 0.74 to 0.97 million in the other four of the six further runs, and the thinking stage took 11,968\,s and 11,704\,s against 4,622 to 6,043\,s in those four. With two fast tries, 62 and 58 exercises reach the thinking tries (24 and 23 in the 4+4 fresh runs), and in a fresh conversation each of them starts its reasoning without the earlier turns; this report did not test which of these features lengthens the thinking stage. The two lanes ran side by side, so these wall clocks are not strictly comparable with runs that had the machine to themselves. On this benchmark, 2+4 fresh gave no accuracy gain over 4+4 fresh.

\subsection{A fresh conversation at the switch}
\label{sec:fresh}
The fast tries are configured identically with and without a fresh conversation, so the choice can matter only for the reasoning tries. In the two corrected 4+4 fresh runs the thinking tries rescued 21 of the 24 exercises still open after four fast tries and 17 of 23, against 15 of 24, 16 of 26 and 16 of 24 in the three 4+4 keep runs (Table~\ref{tab:rescued}). The corrected fresh runs solved 222 and 219 against 215 to 217 for keep, and all six paired comparisons of the two corrected fresh runs with the three keep runs favour the fresh run (Table~\ref{tab:freshfour}: 8 against 2, 8 against 1 and 6 against 1 for run 1; 5 against 2, 5 against 1 and 4 against 2 for run 2).

We read this as a likely modest benefit of a fresh conversation, not established in size. The direction is consistent across runs, but each comparison is small; the six comparisons are not independent, because both fresh runs are compared with the same three keep runs; and the two corrected fresh runs differ from each other by three exercises (222 and 219, paired 4 against 1).

With 2+4 the evidence is weaker. The two fresh runs rescued 59 of the 62 and 52 of the 58 exercises still open at the switch, against 51 of 56 and 56 of 65 for the two keep runs. Three of the four paired comparisons favour the fresh run, and one (the second fresh run against the first keep run, 3 against 4) favours keep (Table~\ref{tab:freshtwo}).

The provisional 4+4 fresh run, whose hand-off note wrongly attributed the earlier attempts to another model, solved 220, with its thinking tries rescuing 16 of 21; paired against the three 4+4 keep runs it gave 5 against 1, 6 against 1 and 4 against 1. It is shown for completeness and is not counted among the six comparisons above or pooled with the corrected runs.

\begin{table}[t]
\centering
\caption{Exercises still open at the switch from fast to thinking tries (out of 225) and how many the thinking tries then solved. For the matched baseline the ``switch'' is after try 4 and the further solves come from six more fast tries. The provisional run is set apart.}
\label{tab:rescued}
\footnotesize
\begin{tabular}{@{}llcc@{}}
\toprule
Run & Conversation & Open at switch & Rescued by thinking tries \\
\midrule
Matched baseline, 10 fast & --- & 27 & 7 (by fast tries) \\
\midrule
4+4, first run & keep & 24 & 15 \\
4+4, repeat & keep & 26 & 16 \\
4+4, 300\,s cap (inert) & keep & 24 & 16 \\
4+4, run 1 (corrected note) & fresh & 24 & 21 \\
4+4, run 2 (corrected note) & fresh & 23 & 17 \\
3+4, first run & keep & 36 & 30 \\
3+4, second run & keep & 33 & 29 \\
2+4, first run & keep & 56 & 51 \\
2+4, second run & keep & 65 & 56 \\
2+4, run 1 & fresh & 62 & 59 \\
2+4, run 2 & fresh & 58 & 52 \\
\midrule
4+4, provisional (note named another model) & fresh & 21 & 16 \\
\bottomrule
\end{tabular}
\end{table}

\begin{table}[t]
\centering
\caption{4+4, fresh against keep: exercises solved only by the fresh run / only by the keep run, against each of the three 4+4 keep runs. The provisional run is set apart and is not pooled.}
\label{tab:freshfour}
\footnotesize
\begin{tabular}{@{}lccc@{}}
\toprule
Fresh run (solved) & vs first keep run (216) & vs repeat (215) & vs 300\,s cap run (217) \\
\midrule
Run 1 (222) & 8 / 2 & 8 / 1 & 6 / 1 \\
Run 2 (219) & 5 / 2 & 5 / 1 & 4 / 2 \\
\midrule
Provisional (220) & 5 / 1 & 6 / 1 & 4 / 1 \\
\bottomrule
\end{tabular}
\end{table}

\begin{table}[t]
\centering
\caption{2+4, fresh against keep: exercises solved only by the fresh run / only by the keep run.}
\label{tab:freshtwo}
\footnotesize
\begin{tabular}{@{}lcc@{}}
\toprule
Fresh run (solved) & vs first keep run (220) & vs second keep run (216) \\
\midrule
Run 1 (222) & 3 / 1 & 6 / 0 \\
Run 2 (219) & 3 / 4 & 5 / 2 \\
\bottomrule
\end{tabular}
\end{table}

\subsection{The time cap, speed and exercises never solved}
\label{sec:timecap}
\paragraph{The time cap.}
No thinking try reached a time limit in any run. The 300\,s cap on thinking requests never fired: no thinking try reached it and no model timeout was recorded. The run is therefore a third repeat of 4+4, and the cap itself is untested. Its shorter wall clock (2\,h 17\,min) cannot be attributed to the cap; it ran concurrently with the 3+4 run, and wall clocks of runs that ran side by side are not strictly comparable. In the six further runs the thinking tries had the default limit of 1800\,s, and none reached it either. A time cap therefore remains untested.

\paragraph{Speed.}
The escalation runs solve more exercises, but their median time to a passing solution is not lower than the baseline's. The single-model runs have medians of 10.9 to 14.8\,s against 10.3\,s for the matched baseline, which has the lowest median and the shortest wall clock (2\,h 07\,min against 2\,h 17\,min to 5\,h 00\,min). Some escalation runs take much longer overall, notably the two 2+4 fresh runs (Section~\ref{sec:earlier}).

\paragraph{Exercises never solved.}
Across all fifteen harness runs (the matched baseline, twelve single-model escalation runs and the two runs of the two-model cascade), cpp/spiral-matrix is the only exercise that was never solved. java/rational-numbers was solved once, in the first 4+4 fresh run with the corrected note. go/robot-simulator, unsolved in the two runs of the two-model cascade and the baseline, was solved twice, in the first 2+4 keep run and in the same 4+4 fresh run. The unsolved lists of the earlier variants contain 10, 5, 6, 8 and 5 exercises (4+4 repeat, 2+4 first run, 3+4 first run, capped run, provisional fresh run) and those of the six further runs 4, 9, 3, 6, 3 and 6 (3+4 second run, 2+4 second run, 4+4 fresh runs 1 and 2, 2+4 fresh runs 1 and 2). cpp/meetup appears in five of these eleven lists.

\subsection{Adding a second model: the two-model cascade}
\label{sec:cascade}
The two-model cascade (Table~\ref{tab:cascade}) differs from 4+4 only in who makes the last two tries: a second model on a second GPU. It was run twice unchanged and is the ablation of adding a second model to the single-model schedules above. This subsection keeps its tables together: the comparison with the matched baseline, the per-try yield, where the gain and the wall clock went, the replication, the matched ablation, and the comparison with the single-model runs.

\subsubsection{Main comparison}
Table~\ref{tab:main} compares the two-model cascade with the matched baseline, which shares its settings; the single-model schedules are in Section~\ref{sec:variants}. The two-model cascade solved 219 of 225 exercises (97.3\%) in at most eight tries, in both runs; the matched baseline solved 205 (91.1\%) in at most ten. In run 1 of the two-model cascade, of all 225 exercises, 39.1\% were solved within 10\,s, 73.8\% within 30\,s and 96.4\% within 5\,min. Among solved exercises, time to solve in run 1 has a mean of 32.1\,s and a 90th percentile of 81.6\,s. There were no infrastructure errors and no timeouts. The matched baseline has the lowest median time to a passing solution (10.3\,s); the two-model cascade does not improve it. The gain appears only in later tries, at the cost of a longer tail and a longer run.

\begin{table}[t]
\centering
\caption{Main comparison of the two-model cascade with the matched baseline (Flash-Next alone, ten tries, summaries off, one thread). ``Solved within 2 tries'' is the leaderboard-style score.}
\label{tab:main}
\footnotesize
\setlength{\tabcolsep}{4pt}
\begin{tabular}{@{}lccc@{}}
\toprule
Measure & \multicolumn{2}{c}{Two-model cascade} & Matched \\
\cmidrule(lr){2-3}
 & run 1 & run 2 & baseline \\
 & (8 tries) & (8 tries) & (10 tries) \\
\midrule
Solved & 219 (97.3\%) & 219 (97.3\%) & 205 (91.1\%) \\
Within 2 tries & 169 (75.1\%) & 161 (71.6\%) & 173 (76.9\%) \\
Median time to solve & 11.5\,s & 12.8\,s & 10.3\,s \\
Wall clock & 3\,h 39\,min & 3\,h 26\,min & 2\,h 07\,min \\
\bottomrule
\end{tabular}
\end{table}

\subsubsection{Per-try yield}
\begin{figure}[t]
\centering
\includegraphics[width=0.65\textwidth]{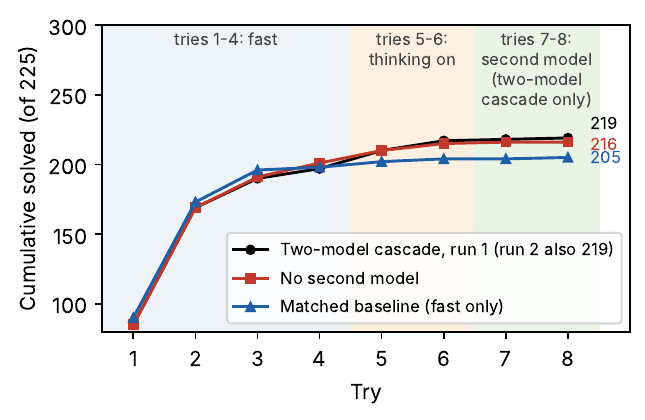}
\caption{Cumulative exercises solved after each try, from the per-try counts in Tables~\ref{tab:yield} and~\ref{tab:atry}: the two-model cascade (run 1), the first run with no second model and the matched baseline. The other single-model runs (Section~\ref{sec:variants}) are not drawn. Shaded bands mark the stages of the two-model cascade; the second-model stage applies to the two-model cascade only. The vertical axis starts at 80, not at zero.}
\label{fig:yield}
\end{figure}

Table~\ref{tab:yield} and Figure~\ref{fig:yield} show the yield of each try in run 1 of the two-model cascade; the repeat is in Section~\ref{sec:replication}. ``Started'' is the number of exercises still unsolved when that try began. It is derived from the per-try solve counts and assumes that every unsolved exercise went on to the next try, which holds for a run without errors. Run 1 used 514 attempts in total: 456 fast, 43 Flash-Next thinking and 15 DeepSeek, or 2.28 attempts per exercise. Only 58 attempts (11\%) used reasoning.

\begin{table}[t]
\centering
\caption{Per-try yield of the two-model cascade, run 1.}
\label{tab:yield}
\begin{tabular}{@{}clrrrr@{}}
\toprule
Try & Stage & Started & Solved & Attempt success & Cumulative solved \\
\midrule
1 & Flash-Next, fast & 225 & 85 & 37.8\% & 85 (37.8\%) \\
2 & Flash-Next, fast & 140 & 84 & 60.0\% & 169 (75.1\%) \\
3 & Flash-Next, fast & 56 & 21 & 37.5\% & 190 (84.4\%) \\
4 & Flash-Next, fast & 35 & 7 & 20.0\% & 197 (87.6\%) \\
5 & Flash-Next, thinking & 28 & 13 & 46.4\% & 210 (93.3\%) \\
6 & Flash-Next, thinking & 15 & 7 & 46.7\% & 217 (96.4\%) \\
7 & DeepSeek, thinking & 8 & 1 & 12.5\% & 218 (96.9\%) \\
8 & DeepSeek, thinking & 7 & 1 & 14.3\% & 219 (97.3\%) \\
\bottomrule
\end{tabular}
\end{table}

In run 1 of the two-model cascade, the success rate of fast retries falls from 60.0\% on try 2 to 20.0\% on try 4. Switching the same model to reasoning mode on try 5 brings it back to 46.4\%. Reasoning mode accounts for 20 of the 22 solves after try 4; the second model accounts for 2. In the repeat, the two reasoning tries solved 14 of the 25 exercises still open after four fast tries, and 14 of the 19 solves after try 4.

\subsubsection{Where the gain comes from}
\label{sec:gain}
Table~\ref{tab:four} contrasts what two kinds of extra effort bought after four fast tries, in two runs with the same settings. In the matched baseline, 198 exercises were solved after four fast tries, and six more fast tries added 7 of the 27 still open. In run 1 of the two-model cascade, 197 were solved after four tries, and two thinking tries added 20 out of 28 (in the repeat, 14 out of 25). In the two-model cascade the thinking tries always ran before the second model, so the effects of reasoning mode and of the second model are confounded by order.

\begin{table}[t]
\centering
\caption{Further solves after four fast tries, in the matched baseline and in run 1 of the two-model cascade.}
\label{tab:four}
\begin{tabular}{@{}llr@{}}
\toprule
After four fast tries & Next step & Further solves \\
\midrule
198 solved, 27 open (matched baseline) & six more fast tries & 7 \\
197 solved, 28 open (two-model cascade) & two thinking tries, same model & 20 \\
\bottomrule
\end{tabular}
\end{table}

The second model's contribution is small. In run 1, eight exercises reached DeepSeek and it solved two; in the repeat it solved five. Those exercises had already survived six attempts by Flash-Next, so they are the hardest in the set, and these counts say little about DeepSeek in general. Whether a second model adds anything beyond reasoning mode is addressed by the ablation and by the single-model runs in Section~\ref{sec:variants}, and the data neither show nor exclude a small benefit.

\subsubsection{Where the wall clock went}
In run 1 of the two-model cascade, the 219 solved exercises account for about 117 minutes (219 $\times$ 32.1\,s). The six unsolved exercises took 78 minutes of model and test time between them (Table~\ref{tab:unsolved}). The rest of the 219-minute run is harness overhead between exercises. In run 1, about a third of the run went to six exercises that were never solved, and four of them account for nearly all of it. A time budget on the reasoning stages is the obvious next control. A 300\,s cap on reasoning requests, tried once in a single-model run, never fired and therefore remains untested (Section~\ref{sec:timecap}).

\begin{table}[t]
\centering
\caption{Unsolved exercises in run 1 of the two-model cascade and the time spent on them across all eight tries. In the repeat, java/hangman and java/zipper were solved.}
\label{tab:unsolved}
\begin{tabular}{@{}ll@{}}
\toprule
Unsolved exercise & Time across all 8 tries \\
\midrule
go/robot-simulator & 27\,min 12\,s (4 test timeouts) \\
java/hangman & 19\,min 20\,s \\
java/rational-numbers & 14\,min 26\,s \\
java/zipper & 14\,min 24\,s \\
cpp/kindergarten-garden & 1\,min 40\,s \\
cpp/spiral-matrix & 1\,min 00\,s \\
\bottomrule
\end{tabular}
\end{table}

\subsubsection{Replication}
\label{sec:replication}
The two-model cascade was run a second time with nothing changed: same configuration, harness commit, models and servers. The benchmark container image was rebuilt from the same build file after a routine cleanup removed it. Table~\ref{tab:rep} compares the two runs.

\begin{table}[t]
\centering
\caption{The two-model cascade, run twice unchanged.}
\label{tab:rep}
\begin{tabular}{@{}lcc@{}}
\toprule
Measure & Run 1 & Run 2 (repeat) \\
\midrule
Solved & 219/225 (97.3\%) & 219/225 (97.3\%) \\
Solved within 2 tries & 169 (75.1\%) & 161 (71.6\%) \\
Median time to solve & 11.5\,s & 12.8\,s \\
Mean / 90th percentile & 32.1\,s / 81.6\,s & 39.8\,s / 56.9\,s \\
Wall clock & 3\,h 39\,min & 3\,h 26\,min \\
Solved by Flash-Next, fast (tries 1--4) & 197 & 200 \\
Solved by Flash-Next, thinking (tries 5--6) & 20 & 14 \\
Solved by DeepSeek (tries 7--8) & 2 & 5 \\
Infrastructure errors & 0 & 0 \\
\bottomrule
\end{tabular}
\end{table}

Paired by exercise, 217 exercises were solved in both runs, 2 only in run 1 (cpp/grade-school, rust/xorcism), 2 only in run 2 (java/hangman, java/zipper), and 4 in neither (cpp/kindergarten-garden, cpp/spiral-matrix, go/robot-simulator, java/rational-numbers). The exact McNemar test gives $p = 1.0$. Across the two runs, 221 of the 225 exercises were solved at least once. Table~\ref{tab:yield2} gives the per-try yield of the repeat.

\begin{table}[t]
\centering
\caption{Per-try yield of the two-model cascade, run 2 (repeat).}
\label{tab:yield2}
\begin{tabular}{@{}clrrr@{}}
\toprule
Try & Stage & Started & Solved & Attempt success \\
\midrule
1 & Flash-Next, fast & 225 & 83 & 36.9\% \\
2 & Flash-Next, fast & 142 & 78 & 54.9\% \\
3 & Flash-Next, fast & 64 & 31 & 48.4\% \\
4 & Flash-Next, fast & 33 & 8 & 24.2\% \\
5 & Flash-Next, thinking & 25 & 12 & 48.0\% \\
6 & Flash-Next, thinking & 13 & 2 & 15.4\% \\
7 & DeepSeek, thinking & 11 & 3 & 27.3\% \\
8 & DeepSeek, thinking & 8 & 2 & 25.0\% \\
\bottomrule
\end{tabular}
\end{table}

What held and what did not:
\begin{itemize}
\item \emph{The headline held.} The same number of exercises was solved, and only 4 of 225 final outcomes differed between runs. The final result is more stable than the first-two-try score, which moved by 3.5 points (75.1\% to 71.6\%); a later identical pair of single-model runs differed on nine outcomes (Section~\ref{sec:variants}).
\item \emph{The decay of fast retries held.} In the two runs of the two-model cascade the per-attempt success of fast retries decays to 20--24\% by try 4 and the first reasoning try restores it to 46--48\% (29\% on try 4 and 38\% on try 5 in the no-second-model run).
\item \emph{The second thinking try did not hold.} It solved 7 of 15 in run 1 and 2 of 13 in the repeat. The first thinking try carries most of the benefit.
\item \emph{DeepSeek's share rose from 2 to 5.} Thinking mode plus the second model solved 22 in run 1 and 19 in the repeat, but the split between them moved. The evidence for a separate benefit from the second model is thin; the ablation in the next subsection addresses it.
\end{itemize}

\subsubsection{Matched baseline and ablation}
\label{sec:matched}
Two further runs used the two-model cascade's exact settings (chat-history summaries off, one thread, same harness, same Basalt server). The \emph{matched baseline} is Flash-Next alone, thinking off, up to ten tries. The \emph{no-second-model} run is Flash-Next with four fast tries and then four thinking tries; DeepSeek is not used and only one GPU is needed. Table~\ref{tab:matched} compares all four runs and Table~\ref{tab:atry} gives solves at each try.

\begin{table}[t]
\centering
\caption{Two-model cascade, ablation and matched baseline, all with summaries off and one thread.}
\label{tab:matched}
\footnotesize
\setlength{\tabcolsep}{4pt}
\begin{tabular}{@{}lcccc@{}}
\toprule
Measure & Two-model & Two-model & No second model & Matched baseline \\
 & cascade, run 1 & cascade, run 2 & (4 fast + 4 thinking) & (10 fast) \\
\midrule
Solved & 219 (97.3\%) & 219 (97.3\%) & 216 (96.0\%) & 205 (91.1\%) \\
Solved within 2 tries & 75.1\% & 71.6\% & 75.1\% & 173 (76.9\%) \\
Median time to solve & 11.5\,s & 12.8\,s & 10.9\,s & 10.3\,s \\
Mean / 90th percentile & 32.1 / 81.6\,s & 39.8 / 56.9\,s & 27.0 / 68.4\,s & 17.5 / 37.0\,s \\
Wall clock & 3\,h 39\,min & 3\,h 26\,min & 2\,h 33\,min & 2\,h 07\,min \\
GPUs used & 2 & 2 & 1 & 1 \\
\bottomrule
\end{tabular}
\end{table}

\begin{table}[t]
\centering
\caption{Exercises solved at each try.}
\label{tab:atry}
\small
\begin{tabular}{@{}lcccccccccc@{}}
\toprule
Run & 1 & 2 & 3 & 4 & 5 & 6 & 7 & 8 & 9--10 \\
\midrule
Matched baseline (all fast) & 90 & 83 & 23 & 2 & 4 & 2 & 0 & 1 & 0 \\
No second model (thinking from try 5) & 85 & 84 & 22 & 10 & 9 & 5 & 1 & 0 & not run \\
\bottomrule
\end{tabular}
\end{table}

Table~\ref{tab:mcnemar} gives the paired per-exercise outcomes, with an exact McNemar test on the discordant pairs.

\begin{table}[t]
\centering
\caption{Paired per-exercise outcomes (exact McNemar test on the discordant pairs). Single runs, except that the two-model cascade has two; ``no second model'' is the first 4+4 run, later single-model runs are in Table~\ref{tab:vpaired}. The $p$-values are nominal; see Section~\ref{sec:variance}.}
\label{tab:mcnemar}
\footnotesize
\setlength{\tabcolsep}{4pt}
\begin{tabular}{@{}lccc@{}}
\toprule
Comparison & Solved only by first & Solved only by second & $p$ \\
\midrule
Two-model cascade run 1 vs matched baseline & 14 & 0 & 0.0001 \\
Two-model cascade run 2 vs matched baseline & 15 & 1 & 0.0005 \\
No second model vs matched baseline & 15 & 4 & 0.019 \\
Two-model cascade run 1 vs no second model & 5 & 2 & 0.45 \\
Two-model cascade run 2 vs no second model & 6 & 3 & 0.51 \\
\bottomrule
\end{tabular}
\end{table}

Total attempts, from the solved-at-each-try tables (exercises still open at try $n$ summed over tries): two-model cascade run 1, 514; two-model cascade run 2, 521; no second model, 513; matched baseline, 573. These are counts of attempts, not of cost, since attempts differ greatly in length.

What this settles and what it does not:
\begin{itemize}
\item \emph{The gain survives matched settings.} With identical settings the matched baseline solved 205 and the two-model cascade 219 in both runs (the baseline had two more tries). In the no-second-model run, 201 were solved after four fast tries, and four thinking tries solved 15 of the remaining 24.
\item \emph{Most of the gain needs only one model.} Switching Flash-Next to thinking mode, with no second model and one GPU, reached 216. The two runs of the two-model cascade reached 219. The paired difference between them (5 vs 2, and 6 vs 3) is small beside the 4 outcomes that changed between two identical runs of the two-model cascade and the 9 that changed between two identical single-model runs (Section~\ref{sec:variants}); the other pairs of identical single-model runs are in Table~\ref{tab:repeats}.
\item \emph{A second model is neither shown nor excluded.} Both runs of the two-model cascade finished 3 above the first single-model run, and DeepSeek solved 2 and 5 exercises that had survived six attempts. Single-model runs scored 215 to 222, a range that contains the two-model cascade's 219 (Section~\ref{sec:variants}). The data neither show nor exclude a small benefit from the second model. It costs a second GPU and about an hour per run.
\end{itemize}

\subsubsection{Single-model runs against the two-model cascade}
\label{sec:secondmodel}
A second model was not needed to reach the solve counts we observed: single-model runs solved 215 to 222, and the two-model cascade solved 219 in each of two runs. The cleanest comparison is with 4+4 keep, which differs from the two-model cascade only in who makes tries 7 and 8: 215 to 217 against 219 and 219, a difference of 2 to 4 exercises from two runs of the two-model cascade, which is not established as larger than run-to-run noise; the 2+4 and 3+4 runs differ from the two-model cascade in schedule as well. Paired against the two runs of the two-model cascade, the first 2+4 keep run solved 4 exercises the two-model cascade did not and the two-model cascade 3 it did not, in each of the two runs; the provisional fresh run gave 1 against 0 (run 1) and 3 against 2 (run 2). Paired counts of the six further runs against the two-model cascade runs were not computed. The second model was tested and was not needed to reach the solve counts observed on this benchmark; the data neither show nor exclude a small benefit from it.

\section{Variance and statistical considerations}
\label{sec:variance}
The exercises are the same in every run, so the appropriate test is paired: an exact McNemar test on per-exercise outcomes. Against the matched baseline, the two-model cascade's advantage is clear (Table~\ref{tab:mcnemar}): run 1 solved 14 exercises the baseline did not and the baseline solved none the two-model cascade did not ($p = 0.0001$); run 2 gave 15 against 1 ($p = 0.0005$). The no-second-model run also beat the baseline (15 against 4, $p = 0.019$). The comparisons of the two-model cascade runs with the no-second-model run show no difference the data can resolve (5 against 2, $p = 0.45$; 6 against 3, $p = 0.51$), and they do not exclude a small benefit from the second model. Two identical runs of the two-model cascade differed on 4 of 225 outcomes ($p = 1.0$), and two identical single-model 4+4 runs on 9 (4 against 5); the further pairs of identical runs are in Table~\ref{tab:repeats}, and run-to-run variation is estimated from these pairs only. All eleven further single-model runs also beat the baseline in paired counts (Table~\ref{tab:vpaired}), and differences of a few exercises between variants are not established. The baseline has one run, the two-model cascade two, 4+4 keep three (one with an inert cap), 4+4 fresh two corrected runs (and one provisional run), 2+4 keep two, 2+4 fresh two and 3+4 keep two. The $p$-values are nominal: they are not corrected for the multiple comparisons in Tables~\ref{tab:mcnemar} and~\ref{tab:vpaired} (the first row of Table~\ref{tab:vpaired} repeats the no-second-model row of Table~\ref{tab:mcnemar}), and they treat exercises as independent. No $p$-values were computed for the six further runs, and the paired comparisons of fresh with keep runs share runs and are not independent of each other.

Three independent ten-try runs of Flash-Next with thinking off, on three engines, solved 201 (vLLM, 4-bit), 202 (Strata, 8-bit) and 204 (Basalt, 8-bit) exercises, a spread of 1.3 points (3 of 225). These runs had chat-history summaries on, so they are not settings-matched to the escalation runs and are not used as baselines (Section~\ref{sec:limits}).

Aggregate scores move by several points: the first-two-try score ranged from 71.1\% to 79.6\% across runs configured identically for the first two tries (Tables~\ref{tab:variants} and~\ref{tab:variantsfresh}). Individual exercises move more: across six earlier two-try runs of Flash-Next on three engines, about a third of exercises changed between solved and unsolved from one run to the next, and outputs differed between runs even at temperature 0. Per-language counts move in the same way: for the same model on two engines, Rust was 17 of 30 in one two-try run and 20 of 30 in the other, and Python was 27 and 31 of 34. Early evidence that one model was better at Rust and another at Python did not survive repeat runs, which is why the schedules escalate on failure and do not route by language.

\section{Negative results}
\label{sec:negative}
The runs in this section used earlier settings (chat-history summaries on, up to four threads, different try counts) and are not comparable with Section~\ref{sec:results}. Table~\ref{tab:neg} lists approaches that did not help on this hardware. For scale, a median 13.6\,s solve (in that earlier run) on Basalt breaks down as 5.8\,s reading the prompt, 6.0\,s writing the answer and about 2\,s running tests. Prompt reading is as large as generation, which is why prompt reuse is the next engine change.

\begin{table}[t]
\centering
\caption{What did not work.}
\label{tab:neg}
\small
\begin{tabular}{@{}p{1.9in}p{1.9in}p{2.4in}@{}}
\toprule
Attempt & Result & Reading \\
\midrule
Larger model: Qwen3.5-397B-A17B Q4\_K\_M, llama.cpp, 18 layers' experts in RAM & 84.4\% on 2 tries, median 6.2\,min, 13.3\,h run, about 18.5 tok/s & About 68\,GB of expert weights ran from system RAM with the GPUs near 10\% busy. Accuracy without speed. \\
Reasoning always on: Flash-Next on Strata & 85.8\% on 2 tries, median 2.5\,min, 14\,h 22\,min & Plus 11.6 points on the 2-try score for about 8$\times$ the run time. The fast retry run reached 89.8\% in 3\,h 24\,min. \\
Four concurrent requests: Basalt \texttt{--parallel 4} & 71.1\%, median 0.9\,min versus 0.2\,min, run 1\,h 46\,min versus 1\,h 40\,min & Each request fell to about 40 tok/s (about 135 total) against 240 to 330 for a single request with speculative decoding. No throughput gain. \\
Prompt caching: stock cache flags, then a longest-common-prefix fork & 0.9\% and 1.1\% of prompt tokens reused & Summary and coder prompts share 3 tokens, and Basalt drops checkpoints when a different conversation arrives. With summaries off, reuse rose to 16\%. An offline replay estimates 41 to 47\% is reachable with checkpoints at turn boundaries. \\
Routing by language & Dropped & The per-language differences were run-to-run variation. \\
\bottomrule
\end{tabular}
\end{table}

\section{Comparison with published results}
\label{sec:published}
On the two-try rule the two-model cascade scores 75.1\% in run 1 (71.6\% in run 2, and 71.1\% to 79.6\% across all the runs in this report). The official Aider leaderboard \citep{aider2025leaderboard} tops out at GPT-5 (high) with 88.0\%. The highest-scoring open-weight entry we could identify is DeepSeek-V3.2-Exp (Reasoner) at 74.2\%; the leaderboard does not label entries by licence. Differences of this size are within the 71.1\% to 79.6\% spread we observed across runs configured identically for the first two tries, and our settings differ from the leaderboard's, so no ranking is implied. The page was last refreshed in November 2025, so newer models are absent. The solve rates of 95.6\% to 98.7\% for the escalation runs (97.3\% for the two-model cascade) are work-until-correct numbers, in-sample, on a public benchmark, and are not comparable with leaderboard scores.

The nearer comparison is with systems that iterate until tests pass. The highest such published result we found is the Refact.ai agent on Claude 3.7 Sonnet \citep{refact2025}: 93.3\% with thinking and 92.9\% without. That is a different harness (an agent with a step cap, not aider with a try cap) on a hosted frontier model, so it is a reference point and not a like-for-like comparison. We found nothing published at or above 97\%, and the search was not exhaustive. No published work-until-correct result for GPT-5 was found, and no comparison is made.

\section{Limitations and threats to validity}
\label{sec:limits}
\paragraph{Confounds and statistical limits.}
\begin{itemize}
\item \emph{Few runs.} There are two runs of the two-model cascade, three of 4+4 keep (one with an inert time cap), two each of 2+4 keep, 2+4 fresh and 3+4 keep, two corrected runs of 4+4 fresh (plus one provisional run) and one matched baseline. Differences of a few exercises cannot be resolved.
\item \emph{In-sample, public benchmark, work-until-correct.} All solve rates are in-sample, from a public benchmark and work-until-correct; they are not comparable with two-try leaderboard scores (Section~\ref{sec:published}).
\item \emph{Order confound for the second model.} The second model always runs last, after the reasoning tries, so any benefit it has is measured only on the exercises that survived six attempts.
\item \emph{The order was tuned on this benchmark.} The stage boundaries (4, 2, 2) were chosen after looking at single-model results on the same 225 exercises, and the single-model variants (2, 3 and 4 fast tries) were then compared on the same exercises. There is no held-out set, so the high solve rates are in-sample.
\item \emph{Second-model benefit unresolved.} Both runs of the two-model cascade finished 3 above the first no-second-model run, while single-model runs scored 215 to 222. The data neither show nor exclude a small benefit from the second model.
\item \emph{Try caps differ.} The two-model cascade has 8 tries, the single-model schedules 6 to 8 and the baselines 10. This favours the baselines.
\item \emph{Untested and provisional variants.} The 300\,s thinking cap never fired and no thinking try reached any time limit in any run, so a time cap is untested. One 4+4 fresh run used a hand-off note that wrongly described the earlier attempts as another model's; it is provisional, is reported separately and is not pooled with the two corrected runs. At 4+4 the benefit of a fresh conversation is consistent in direction but not established in size, and its paired comparisons share runs and are not independent.
\item \emph{Concurrent runs and harness version.} Two model instances ran side by side for the single-model variants, so their wall clocks are not strictly comparable. The harness version changed from 0.2.2 to 0.3.0 between the earlier runs and the five earlier variants, and to 0.3.1 for the six further runs, which differs from 0.3.0 only in the wording of the fresh-conversation hand-off brief.
\item \emph{Earlier runs with chat-history summaries on (settings differ).} Two earlier ten-try runs (204 solved for Flash-Next alone, 207 for DeepSeek alone) are not settings-matched to the escalation runs and are not used as baselines. They ran with aider's chat-history summaries on; the two-model cascade and the matched baseline ran with them off. With the benchmark's default (\texttt{max\_chat\_history\_tokens} = 1024), aider sends a summary request after nearly every coder reply, about half of all requests and 26\% of prompt tokens. Turning them off, in a paired test on the same engine, cut wall clock by 29\% (1\,h 54\,min to 1\,h 21\,min) and median time to solve by 27\% (13.9\,s to 10.1\,s) with the same solve rate (74.2\% both). The DeepSeek run also had both GPUs and four sessions, so its times are measured under load.
\end{itemize}

\paragraph{Other limits.}
\begin{itemize}
\item \emph{Scope.} The exercises are small, self-contained and come with complete test suites. The results say nothing about large repositories, ambiguous requirements or tasks without tests.
\item \emph{Possible training-data contamination.} The benchmark is public and old enough that some exercises may be in the training data of both models. This may affect absolute scores more than the comparison between schedules, but we did not test this.
\item \emph{Unequal attempt cost.} A try count is a poor unit, because attempts differ greatly in cost. Wall-clock time to a passing solution is the measure reported here, and the one we would ask others to report.
\item \emph{Harness.} The harness used for these runs is not yet public.
\end{itemize}

\section{Future work}
\label{sec:future}
The following remain open.
\begin{enumerate}
\item A measured time cap on the reasoning stages, to recover the wall clock spent on unsolved exercises; no thinking try has reached any time limit so far, so thinking-try durations should be measured before a cap is chosen.
\item More repeats of each schedule, in particular of 4+4 keep and 4+4 fresh, to size the effect of a fresh conversation at the switch, and of the two-model cascade, to tell whether the second model adds anything.
\item Harder task sets, on which the escalation stages would be exercised more often.
\item A held-out evaluation, with schedule boundaries fixed before the held-out exercises are run, to replace the in-sample figures.
\item Other line-ups and other hardware.
\end{enumerate}

\section{Conclusion}
After repeated fast failures, switching the same model into reasoning mode recovers most of the remaining exercises. With matched settings, a single model with ten fast tries solved 205 of 225 Aider Polyglot exercises (91.1\%); every escalation run solved 215 to 222. The same model with four fast and four reasoning tries on one GPU solved 216, 215 and 217 in three runs (95.6\% to 96.4\%), and the single-model variants that escalate after three or two fast tries or start a fresh conversation at the switch solved 216 to 222. A two-model cascade solved 219 in each of two runs. A second model was not needed to reach the solve counts we observed: single-model runs solved 215 to 222, and the two-model cascade solved 219 in each of two runs; the data neither show nor exclude a small benefit from a second model. At 4+4 a fresh conversation at the switch is consistent in direction (all six paired comparisons favour it) and a likely modest benefit, not established in size; at 2+4 one of four comparisons goes against it. The comparisons share runs. Earlier escalation (after two or three fast tries) gave no accuracy gain distinguishable from run-to-run noise and took longer: 2+4 solved 220 and 216 and 3+4 solved 219 and 221, against 216, 215 and 217 for 4+4, and 2+4 with a fresh conversation matched 4+4 with a fresh conversation in accuracy while taking about five hours against about three and a half. The gain is more exercises solved, not a lower median time: the matched baseline has the lowest median time to a passing solution (10.3\,s against 10.9 to 14.8\,s), and some escalation runs take much longer overall. No thinking try reached a time limit, so a time cap remains untested. Run-to-run noise is a few exercises, so differences of a few exercises between variants are not established.
The evidence rests on few runs per variant (4+4 keep three, 4+4 fresh two corrected runs, 2+4 keep two, 2+4 fresh two, 3+4 keep two, two-model cascade two, baseline one), with tuned stage boundaries on a public benchmark, so the high solve rates are in-sample and are not comparable with two-try leaderboard scores.

\section*{Data availability}
The results data for all runs, including failed and abandoned ones, together with the full tables and descriptions of the engines and models, are public at \url{https://github.com/AI-Development-Work/aimos-benchmark-results}. The harness used to produce them is not yet public.

\section*{Acknowledgment of AI assistance}
Generative AI tools (Anthropic's Claude) were used to assist with running experiments, writing analysis scripts and drafting text. The author reviewed the content and takes full responsibility for it.

\bibliographystyle{plainnat}
{\footnotesize\setlength{\bibsep}{1pt}
\bibliography{refs}}
\end{document}